\documentclass[preprint,11pt]{article}%
\usepackage{amsfonts}
\usepackage{amsmath}
\usepackage{amssymb}
\usepackage{graphicx}%
\title{Discussing Quantum Aspects of Higher-Derivative 3D-Gravity in the First-Order Formalism}
\author{J. A. Helay\"{e}l-Neto $^{\dagger}$, L. M. de Moraes $^{*}$ and V. J. Vasquez Otoya$^{\ddagger}$\\ $^{\dagger }$ CBPF, Centro Brasileiro de Pesquisas F\'{\i}sicas \\
Rua Xavier Sigaud 150, 22290-180, Urca \\ Rio de Janeiro, Brazil\\ $^{\ast}$ Centro Federal de Educa\c{c}\~{a}o Tecnol\'{o}gica Celso Suckow da Fonseca\\Unidade Descentralizada de Nova Friburgo \\Av. Governador Roberto Silveira, 1900. Prado, Nova Friburgo\\ Rio de Janeiro, Brazil\\$^{\ddagger}$UFF, Universidade Federal Fluminense \\
Campus da Praia Vermelha, Gragoata, 24210-310, Niteroi \\
Rio de Janeiro, Brazil}
\begin{document}
\maketitle
\begin{abstract}
In this paper, we reassess the issue of deriving the propagators and
identifying the spectrum of excitations associated to the vielbein and spin
connection of (1+2)-D gravity in the presence of dynamical torsion, while working in the
first-order formulation. A number of peculiarities is pointed out whenever the
Chern-Simons term is taken into account along with a combination of bilinear terms
in the torsion tensor. We present a procedure to derive the full set of
propagators, based on an algebra of enlarged spin-type operators, and we discuss under which
conditions the poles of the tree-level 2-point functions correspond to
physical excitations that do not conflict with causality and unitarity.
\end{abstract}

\section{Introduction}

The need to better understand gauge fields has lead to an widespread use of
local transformations due the natural manner gauge fields appear in it. In the
attempt to write (1+2)-D gravity as a gauge theory, the formulation requires
some specific technicalities, by virtue of the possibility of including the
so-called (topological) Chern-Simons term. Adopting the Poincar\'{e} group as
the local gauge group, one naturally obtains the curvature and torsion tensor
by means of the Cartan%
\'{}%
s structure equations. The translational part of the Poincar\'{e} group is
represented by the vielbein gauge fields, $e_{\alpha}{}^{\mathsf{a}}$, which
are also diffeomorphic invariant under general coordinate transformations, and
the Lorentzian part --- realizing the equivalence principle --- given by the
spin connection gauge fields, $\omega_{\alpha}{}^{\mathsf{ab}}$. The vielbein
fields associate to each point a locally flat coordinate system and the spin
connection relates any two local Lorentz coordinate systems at the given point.

This formalism is believed to be completely equivalent to the formalism that
employs affine connections and define curvature and torsion in terms of them.
There is a great deal of results that confirm this, mainly at the level of
expressions to the curvature and torsion. At the classical level, this
equivalence is indeed true. However, when we go over to the quantum
field-theoretic version, there appear remarkable differences and we must
indeed adopt the vielbein and spin connection as the independent fundamental
degrees of freedom \cite{Arcos:2005ec}.

More recently, there has been a considerable raise of interest in $(1+2)$-dimensional gravity models with higher powers of the curvature. Specially, the New Massive Gravity Theory proposed by Bergshoeff, Hohm and Townsend \cite{Bergshoeff:2009hq, Bergshoeff:2009aq, Bergshoeff:2009fj} has triggered a number of contributions to the study of remarkable peculiarities and properties of higher-order $3d$-gravity models \cite{Nakasone:2009bn}-\cite{ Andringa:2009yc}.

Motivated by these these recent results, we propose here to pursue an investigation of planar gravity actions with propagating tensor, since we have in mind to understand which role torsion degrees of freedom may play in connection with unitary higher-derivative gravitation in $3D$. In order to investigate this further, we begin with the analysis of a traditional (1+2)-D gravity model previously done by two of us \cite{Boldo:1999qw}, where we have studied the inclusion of torsion in three-dimensional
Einstein-Chern-Simons gravity and added up higher-derivative terms; we have worked in affine connection formalism. In this work, due to invertibility problems that
appear in the theory if we adopt the first-order approach, we are lead to change to another Lagrangian density, where we introduce torsion algebraic terms, yielding dynamical spin-connection.

Due to the importance of the torsion terms, it is worthwhile to remember that
torsion was introduced by E. Cartan in 1922, as the antisymmetric part of the
affine connection and was recognized by him as a geometric object related to
an intrinsic angular moment of matter. After the introduction of the spin
concept, it was suggested that torsion should mediate a contact interaction
between spinning particles without propagation in matter-free space
\cite{Hehl:1976kj, Hehl:1994ue, Mielke:1991nn}. Later,\ due the fact that at the microscopic level
particles are classified by their mass and spin according to the Poincar\'{e}
group, gauge theories of General Relativity were developed that brings in it
dynamic torsion \cite{DeSabbata:1980gb}. These theories are motivated
by the requirement that the Dirac equation in a gravitational field preserves
local invariance under Lorentz transformations which yields, across the
minimal coupling, a direct interaction between torsion and fermions.
Observational constraints for a propagating torsion and its matter
interactions are discussed in \cite{Carroll:1994dq, Hammond:1995rp, Belyaev:1997zv, sabbata, Shapiro:1994vs, Shapiro:2001rz}.

Our work is organised according to the following outline: in Section 2, we
present a quick review of the Einstein-Cartan formalism, with the purpose of
fixing notation and setting our conventions. Next, the general model and the
decomposition of the action in terms of spin operators is the subject of
Section 3, where we point out a serious problem related to a spin-2
excitation. This motivates us to introduce and to analyse a number of torsion
terms in the action, which is done in Section 4. In Section 5, we come to the
task of computing the propagators and we analyse thereby their poles with the
corresponding residues, in order to locate the physically relevant regions in
the parameter space. Finally, in Section 6, we present our Concluding
Comments, with a critical discussion on our main results and possible issues
for future investigation.

\section{Well-known Results on the Einstein-Cartan Approach}

A Riemann-Cartan space-time \cite{Hehl:1976kj, sabbata, Shapiro:1994vs, Shapiro:2001rz}\ is defined as a
manifold where the covariant derivative of the metric field exists and is
given by:%

\begin{equation}
\nabla_{\gamma}g_{\alpha\beta}(x)=0, \label{um}%
\end{equation}
where this equation defines the so called metric-compatible affine connection,
$\Gamma_{\alpha\beta}{}^{\gamma}$; it allows the presence of torsion, given by
the antisymmetric part of the affine connection,%

\begin{equation}
\mathcal{T}_{\alpha\beta}{}^{\gamma}=2\Gamma_{\lbrack\alpha\beta]}{}^{\gamma}.
\label{dois}%
\end{equation}

We then have:%
\begin{equation}
\Gamma_{\alpha\beta}{}^{\gamma}=\left\{  _{\alpha\beta}^{\ \gamma}\right\}
+K_{\alpha\beta}{}^{\gamma}, \label{tres}%
\end{equation}
where $\left\{  _{\alpha\beta}^{\ \gamma}\right\}  $ is the Christoffel
symbol, which is completely determined by the metric,%

\begin{equation}
\left\{  _{\alpha\beta}^{\ \gamma}\right\}  =\frac{1}{2}g^{\gamma\lambda
}(\partial_{\alpha}g_{\lambda\beta}+\partial_{\beta}g_{\alpha\lambda}%
-\partial_{\lambda}g_{\alpha\beta})
\end{equation}
and%

\begin{equation}
K_{\alpha\beta}{}^{\gamma}=\frac{1}{2}(\mathcal{T}_{\alpha\beta}{}^{\gamma
}+\mathcal{T}^{\ \gamma}{}_{\alpha\beta}-\mathcal{T}_{\beta}{}^{\gamma}%
{}_{\alpha})
\end{equation}
is the contortion tensor, antisymmetric in the last two indices.

In order to study local properties one introduces (in our specific (1+2)-D
case) the dreibein vector fields, $e_{\alpha}{}^{\mathsf{a}}(x)$,\ that spans
at any given point the local Minkowski space-time, which in this work has
metric: $\eta_{\mathsf{ab}}=diag(1,-1,-1)$.

The introduction of the tangent Minkowski space-time allows local Lorentz
transformations on geometrical objects (with Latin index). In order to render
these objects invariant under local Lorentz rotations, one introduces the spin
connection $\omega_{\mathsf{\gamma b}}{}^{\mathsf{c}}$. The covariant
derivative of the dreibein then reads:%

\begin{equation}
\nabla_{\gamma}e_{\alpha}{}^{\mathsf{a}}=D_{\gamma}e_{\alpha}{}^{\mathsf{a}%
}-\Gamma_{\gamma\alpha}{}^{\lambda}e_{\lambda}{}^{\mathsf{a}}=0, \label{seis}%
\end{equation}
where $D_{\gamma}e_{\alpha}{}^{\mathsf{a}}=\partial_{\gamma}e_{\alpha}%
{}^{\mathsf{a}}+\omega_{\gamma\mathsf{i}}{}^{\mathsf{a}}e_{\alpha}%
{}^{\mathsf{i}}$ is the Lorentz covariant derivative.

One finds from eq.(\ref{seis}) that the affine connection can then be written as:%

\begin{equation}
\Gamma_{\alpha\beta}{}^{\gamma}=e_{\mathsf{j}}{}^{\gamma}D_{\alpha}e_{\beta}%
{}^{\mathsf{j}}, \label{sete}%
\end{equation}
and the torsion tensor, eq.(\ref{dois}), reads%

\begin{equation}
\mathcal{T}_{\alpha\beta}{}^{\gamma}=2\Gamma_{\lbrack\alpha\beta]}{}^{\gamma
}=e_{\mathsf{j}}{}^{\gamma}(\partial_{\alpha}e_{\beta}{}^{\mathsf{j}}%
-\partial_{\beta}e_{\alpha}{}^{\mathsf{j}}+\omega_{\alpha\mathsf{i}}%
{}^{\mathsf{j}}e_{\beta}{}^{\mathsf{i}}-\omega_{\beta\mathsf{i}}{}%
^{\mathsf{j}}e_{\alpha}{}^{\mathsf{i}}). \label{doze}%
\end{equation}

As known, the curvature tensors and scalar are given in terms of the affine
connection by the expressions:%

\begin{equation}
\mathcal{R}_{\mu\alpha\beta}{}^{\nu}=\partial_{\mu}\Gamma_{\alpha\beta}{}%
^{\nu}-\partial_{\alpha}\Gamma_{\mu\beta}{}^{\nu}+\Gamma_{\mu\rho}{}^{\nu
}\Gamma_{\alpha\beta}{}^{\rho}-\Gamma_{\alpha\rho}{}^{\nu}\Gamma_{\mu\beta}%
{}^{\rho},
\end{equation}

\begin{equation}
\mathcal{R}_{\alpha\beta}=\mathcal{R}_{\mu\alpha\beta}{}^{\mu}=\partial_{\mu
}\Gamma_{\alpha\beta}{}^{\mu}-\partial_{\alpha}\Gamma_{\mu\beta}{}^{\mu
}+\Gamma_{\mu\rho}{}^{\mu}\Gamma_{\alpha\beta}{}^{\rho}-\Gamma_{\alpha\rho}%
{}^{\mu}\Gamma_{\mu\beta}{}^{\rho}%
\end{equation}
and%

\begin{equation}
\mathcal{R}=g^{\alpha\beta}\mathcal{R}_{\alpha\beta}.
\end{equation}

In terms of the spin connection,%

\begin{equation}
\mathcal{R}_{\mu\alpha\beta}{}^{\nu}=e_{\beta}{}^{\mathsf{i}}e_{\mathsf{j}}%
{}^{\nu}(\partial_{\mu}\omega_{\alpha\mathsf{i}}{}^{\mathsf{j}}-\partial
_{\alpha}\omega_{\mu\mathsf{i}}{}^{\mathsf{j}}+\omega_{\mu\mathsf{k}}%
{}^{\mathsf{j}}\omega_{\alpha\mathsf{i}}{}^{\mathsf{k}}-\omega_{\alpha
\mathsf{k}}{}^{\mathsf{j}}\omega_{\mu\mathsf{i}}{}^{\mathsf{k}}),
\end{equation}%
\begin{equation}
\mathcal{R}_{\alpha\beta}=e_{\beta}{}^{\mathsf{i}}e_{\mathsf{j}}{}^{\mu
}(\partial_{\mu}\omega_{\alpha\mathsf{i}}{}^{\mathsf{j}}-\partial_{\alpha
}\omega_{\mu\mathsf{i}}{}^{\mathsf{j}}+\omega_{\mu\mathsf{k}}{}^{\mathsf{j}%
}\omega_{\alpha\mathsf{i}}{}^{\mathsf{k}}-\omega_{\alpha\mathsf{k}}%
{}^{\mathsf{j}}\omega_{\mu\mathsf{i}}{}^{\mathsf{k}}) \label{oito}%
\end{equation}
and%

\begin{equation}
\mathcal{R}=\eta^{\mathsf{ai}}e_{\mathsf{a}}{}^{\alpha}e_{\mathsf{j}}{}^{\mu
}(\partial_{\mu}\omega_{\alpha\mathsf{i}}{}^{\mathsf{j}}-\partial_{\alpha
}\omega_{\mu\mathsf{i}}{}^{\mathsf{j}}+\omega_{\mu\mathsf{k}}{}^{\mathsf{j}%
}\omega_{\alpha\mathsf{i}}{}^{\mathsf{k}}-\omega_{\alpha\mathsf{k}}%
{}^{\mathsf{j}}\omega_{\mu\mathsf{i}}{}^{\mathsf{k}}). \label{nove}%
\end{equation}

\section{A Problem Related to a Spin-2 Excitation}

We start off with the three-dimensional action for topologically massive gravity:%

\begin{equation}
\mathcal{S=}\int d^{3}x\ e\left(  \mathsf{a}_{1}\mathcal{R+}\mathsf{a}%
_{2}\mathcal{R}^{2}+\mathsf{a}_{3}\mathcal{R}_{\alpha\beta}\mathcal{R}%
^{\alpha\beta}+\mathsf{a}_{4}\mathcal{L}_{CS}\right)  , \label{dez}%
\end{equation}
where%

\begin{equation}
\mathcal{L}_{CS}=\varepsilon^{\alpha\beta\gamma}\Gamma_{\gamma\delta}%
{}^{\lambda}\left(  \partial_{\alpha}\Gamma_{\lambda\beta}{}^{\delta}+\frac
{2}{3}\Gamma_{\alpha\rho}{}^{\delta}\Gamma_{\beta\lambda}{}^{\rho}\right)  ,
\label{quatorze}%
\end{equation}
is the topological Chern-Simons term \cite{Deser:1982vy, Deser:1981wh}
 and%

\begin{equation}
\varepsilon^{\alpha\beta\gamma}=\frac{\epsilon^{\alpha\beta\gamma}}{e}%
\end{equation}
is the completely antisymmetric tensor in (1+2)-D, with $\epsilon^{\alpha
\beta\gamma}$ the Levi-Civita tensor density in the flat space and $e=\sqrt
{g}$ where $g=\det(g_{\alpha\beta})=\eta e^{2}$. $\mathsf{a}_{1}$,
$\mathsf{a}_{2}$ and $\mathsf{a}_{3}$ are free coefficients, whereas
$\mathsf{a}_{4}$ is the Chern-Simons parameter. For a clear and detailed discussion of theories with Chern-Simons term, see
\cite{Zanelli:2002qm}.

As the Riemann tensor, $\mathcal{R}_{\mu\alpha\beta}{}^{\nu}$, it has the same
number of independent \ components as the Ricci tensor, $\mathcal{R}%
_{\alpha\beta}$, in three dimensions, a term squared in $\mathcal{R}%
_{\mu\alpha\beta}{}^{\nu}$ is not necessary in the action of eq.(\ref{dez})

In \cite{Boldo:1999qw}, we have written the affine connection as in eq.(\ref{tres}),
further decomposing the torsion in its SO(1,2) irreducible components: a
scalar from the totally antisymmetric part, a three-vector from the trace and
a symmetric traceless rank-2 tensor. With this procedure, we have obtained a
particle spectrum where only massive excitations of spin-2 associated with the
linearized gravitational field, $h^{\alpha\beta}$, and with the symmetric part
of the torsion field had dynamics that preserved the unitarity of the theory
for some values of the action parameters. In view of the results we obtained
there we saw that a ghost-free 3D gravity theory can be formulated once some
constraints are imposed on the parameters of the Lagrangians we dicuss
\cite{Sezgin:1979zf, Sezgin:1981xs}. Thus, the interesting result that arised from
our previous discussion is the possibility to write down a higher-derivative
model for 3-D gravity with unitarity under control.

In this section, we reconsider the action (\ref{dez}) but, contrary to what we
have done in \cite{Boldo:1999qw}, we propose to adopt in the first-order
formulation, dropping the torsion as a fundamental excitation and electing
the dreibein and the spin connection as the fundamental quantum fields, which
reveals the full gauge structure of gravity.

Now, we consider eqs. (\ref{sete}),(\ref{oito}) and (\ref{nove}), and we adopt  the weak field decomposition of the gravitational field,%

\begin{equation}
e_{\alpha}{}^{\mathsf{a}}=\delta_{\alpha}{}^{\mathsf{a}}+\frac{k}{2}H_{\alpha
}{}^{\mathsf{a}}\left(  \Rightarrow g_{\alpha\beta}=\eta_{\alpha\beta
}+kh_{a\beta},\ \ h_{\alpha\beta}=\frac{1}{2}(H_{\alpha\beta}+H_{\beta\alpha
})\right)  , \label{vintedois}%
\end{equation}
where $k$ is the Planck lengh, 

\begin{equation}
H_{\mathsf{ab}}=h_{\mathsf{ab}}+\mathcal{H}_{\mathsf{ab}}\ ,\ h_{\mathsf{ab}%
}=H_{(\mathsf{ab)}}\ \ e\ \ \mathcal{H}_{\mathsf{ab}}=H_{[\mathsf{ab]}}
\label{vintetres}%
\end{equation}
and
\begin{equation}
\mathcal{H}_{\mathsf{ab}}=\epsilon_{\mathsf{abc}}h^{\mathsf{c}}\Rightarrow
h_{\mathsf{a}}=\frac{1}{2}\epsilon_{\mathsf{abc}}\mathcal{H}^{\mathsf{bc}}.
\label{vintequatro}%
\end{equation}

The spin connection can be written in terms of its dual as follows:%

\begin{equation}
\omega_{\mathsf{a}}{}^{\mathsf{bc}}=\epsilon^{\mathsf{bcd}}Y_{\mathsf{ad}},
\label{quinze}%
\end{equation}
which can be further split according to,%

\begin{equation}
Y_{\mathsf{ab}}=y_{\mathsf{ab}}+\mathcal{Y}_{\mathsf{ab}}\ ;\ y_{\mathsf{ab}%
}=Y_{(\mathsf{ab)}}\ \ ,\ \ \mathcal{Y}_{\mathsf{ab}}=Y_{[\mathsf{ab]}}
\label{dezesseis}%
\end{equation}
and%

\begin{equation}
\mathcal{Y}_{\mathsf{ab}}=\epsilon_{\mathsf{abc}}y^{\mathsf{c}}\Rightarrow
y_{\mathsf{a}}=\frac{1}{2}\epsilon_{\mathsf{abc}}\mathcal{Y}^{\mathsf{bc}}.
\label{dezessete}%
\end{equation}
We then  rewrite the action (\ref{dez}), to which we add the gauge-fixing terms,%

\begin{equation}
\mathcal{L}_{GF-diff}=\lambda F_{\mathsf{a}}F^{\mathsf{a}}\ ,\ F_{\mathsf{a}%
}=\partial_{\mathsf{b}}\left(  H_{\mathsf{a}}^{\mathsf{b}}-\frac{1}{2}%
\delta_{\mathsf{a}}^{\mathsf{b}}H_{\mathsf{c}}^{\mathsf{c}}\right)  ,
\end{equation}
and%

\begin{equation}
\mathcal{L}_{GF-LL}=\xi\left(  \partial^{\mu}\omega_{\mu}{}^{\mathsf{ab}%
}\partial^{\nu}\omega_{\nu\mathsf{ab}}\right)  ,
\end{equation}

in a more suitable (linearized) form:%

\begin{equation}
\mathcal{S}\mathcal{=}\int d^{3}x\frac{1}{2}\Phi^{T}M\Phi\ ,\ \Phi=\left(
\begin{array}
[c]{c}%
y^{\mathsf{cd}}\\
y^{\mathsf{c}}\\
h^{\mathsf{cd}}\\
h^{\mathsf{c}}%
\end{array}
\right)  .
\end{equation}
$\lambda$ e $\xi$ are the gauge-fixing parameters. The wave operator, $M$, can be expressed in terms of an extension of the
spin-projection operator formalism introduced in \cite{rivers}%
,\cite{VanNieuwenhuizen:1973fi, Nunes:1993yu, Boldo:1999qw}.

Here, we would like to point out that, if we are concerned just with the
excitation spectrum associated to the model under consideration and its
unitarity property, we could simply decompose the fields according to their
irreducible components, diagonalise the bilinear piece of the action (this
would split the physical field components from the gauge compensating ones)
and them read off the spectrum. However, the field components so obtained are
non-local fields, since a $\square^{-1}$\ appears in the projectors which act
on the fields to separate their physical components. Since we wish to get the
propagators for the local fields, having in mind that we can later carry out
pertubative loop computations, we choose to keep the full fields and we are
then obliged to gauge-fix the action so as to give propagation to the
compensating components, and invert the wave operator, $M$.

In this article, we follow the notations of \cite{VanNieuwenhuizen:1973fi, Nunes:1993yu} for the
Barnes-Rivers operators, where it refers to the energy-momentum tensor: $(2)$
is the pure spin-2 sector, $(1-m)$ is the part related to the spin-1 momentum
vector, $(0-s)$\ is the part related to the spin-0 stress scalar and
$(0-w)$\ is the part related to the spin-0 work (energy) scalar; $(0-sw)$\ and
$(0-ws)$\ are operators that map the spaces with the same spin. Five
additional operators coming from the $y^{\mathsf{a}}$ -- and Chern-Simons
terms are needed, where the notation $(2a)$ indicates that this operator is a
spin-2 operator with comutation relations only with the pure spin-2,\ in
analogy to $(1a)$. Throughout this work, it is supposed that all differential
operators that appear in the spin operators are duely replaced by a momentum
3-vector, in Fourier space.

The six operators for a rank-2 symmetric tensor in 3D are then given by:%

\[
\mathsf{P}_{\mathsf{ab,cd}}^{(2)}=\frac{1}{2}(\theta_{\mathsf{ac}}%
\theta_{\mathsf{bd}}+\theta_{\mathsf{ad}}\theta_{\mathsf{bc}})-\frac{1}%
{2}\theta_{\mathsf{ab}}\theta_{\mathsf{cd}},
\]

\[
\mathsf{P}_{\mathsf{ab,cd}}^{(1-m)}=\frac{1}{2}(\theta_{\mathsf{ac}}%
\omega_{\mathsf{bd}}+\theta_{\mathsf{ad}}\omega_{\mathsf{bc}}+\theta
_{\mathsf{bc}}\omega_{\mathsf{ad}}+\theta_{\mathsf{bc}}\omega_{\mathsf{ad}}),
\]

\begin{equation}
\mathsf{P}_{\mathsf{ab,cd}}^{(0-s)}=\frac{1}{2}\theta_{\mathsf{ab}}%
\theta_{\mathsf{cd}},
\end{equation}

\[
\mathsf{P}_{\mathsf{ab,cd}}^{(0-w)}=\omega_{\mathsf{ab}}\omega_{\mathsf{cd}},
\]

\[
\mathsf{P}_{\mathsf{ab,cd}}^{(0-sw)}=\frac{1}{\sqrt{2}}\theta_{\mathsf{ab}%
}\omega_{\mathsf{cd}}%
\]
and%

\[
\mathsf{P}_{\mathsf{ab,cd}}^{(0-ws)}=\frac{1}{\sqrt{2}}\omega_{\mathsf{ab}%
}\theta_{\mathsf{cd}},
\]
where $\theta_{\mathsf{ab}}=\eta_{\mathsf{ab}}-\omega_{\mathsf{ab}}$ is the
transverse and $\omega_{\mathsf{ab}}=\frac{\partial_{\mathsf{a}}%
\partial_{\mathsf{b}}}{\square}$ is the longitudinal projector operator that
act on vector fields to split their spin-0 and spin-1 components. The others
five operators are:%

\[
\mathsf{S}_{\mathsf{ab,cd}}^{(2a)}=(\epsilon_{\mathsf{ace}}\theta
_{\mathsf{bd}}+\epsilon_{\mathsf{ade}}\theta_{\mathsf{bc}}+\epsilon
_{\mathsf{bce}}\theta_{\mathsf{ad}}+\epsilon_{\mathsf{bce}}\theta
_{\mathsf{ad}})\partial^{\mathsf{e}},
\]

\[
\mathsf{R}_{\mathsf{ab,cd}}^{(1a)}=(\epsilon_{\mathsf{ace}}\omega
_{\mathsf{bd}}+\epsilon_{\mathsf{ade}}\omega_{\mathsf{bc}}+\epsilon
_{\mathsf{bce}}\omega_{\mathsf{ad}}+\epsilon_{\mathsf{bce}}\omega
_{\mathsf{ad}})\partial^{\mathsf{e}},
\]

\begin{equation}
\mathsf{A}_{\mathsf{ab}}=\epsilon_{\mathsf{abc}}\partial^{\mathsf{c}},
\label{operators}%
\end{equation}

\[
\mathsf{B}_{\mathsf{a,bc}}=\eta_{\mathsf{ab}}\partial_{\mathsf{c}}%
+\eta_{\mathsf{ac}}\partial_{\mathsf{b}}%
\]
and%

\[
\mathsf{D}_{\mathsf{a,bc}}=\mathsf{A}_{\mathsf{ab}}\partial_{\mathsf{c}%
}+\mathsf{A}_{\mathsf{ac}}\partial_{\mathsf{b}}.
\]

We recall that the usual Barnes-Rivers operators obey the algebra:%

\[
\mathsf{P}_{\mathsf{ab,kl}}^{(i-a)}\mathsf{P}^{(j-b)\ \mathsf{kl}}%
{}_{,\mathsf{cd}}=\delta^{ij}\delta^{ab}\mathsf{P}_{\mathsf{ab,cd}}^{(j-b)},
\]

\begin{equation}
\mathsf{P}_{\mathsf{ab,kl}}^{(i-ab)}\mathsf{P}^{(j-cd)\ \mathsf{kl}}%
{}_{,\mathsf{cd}}=\delta^{ij}\delta^{bc}\mathsf{P}_{\mathsf{ab,cd}}^{(j-a)},
\end{equation}

\[
\mathsf{P}_{\mathsf{ab,kl}}^{(i-a)}\mathsf{P}^{(j-bc)\ \mathsf{kl}}%
{}_{,\mathsf{cd}}=\delta^{ij}\delta^{ab}\mathsf{P}_{\mathsf{ab,cd}}^{(j-ac)},
\]

\[
\mathsf{P}_{\mathsf{ab,kl}}^{(i-ab)}\mathsf{P}^{(j-c)\ \mathsf{kl}}%
{}_{,\mathsf{cd}}=\delta^{ij}\delta^{bc}\mathsf{P}_{\mathsf{ab,cd}}^{(j-ac)}%
\]
and satisfy the tensor identity,%

\begin{equation}
\mathsf{P}_{\mathsf{ab,cd}}^{(2)}+\mathsf{P}_{\mathsf{ab,cd}}^{(1m)}%
+\mathsf{P}_{\mathsf{ab,cd}}^{(0s)}+\mathsf{P}_{\mathsf{ab,cd}}^{(0w)}%
=\frac{1}{2}\left(  \eta_{\mathsf{ac}}\eta_{\mathsf{bd}}+\eta_{\mathsf{ad}%
}\eta_{\mathsf{bc}}\right)  . \label{onze}%
\end{equation}

The new set of spin operators that comes about displays, besides the operators
$\mathsf{S}_{\mathsf{ab,cd}}^{(2a)}$, $\mathsf{R}_{\mathsf{ab,cd}}^{(1a)}$,
$\mathsf{A}_{\mathsf{ab}}$, and $\mathsf{B}_{\mathsf{a,bc}}$\ (already known
from \cite{Boldo:1999qw}), one new operator, $\mathsf{D}_{\mathsf{a,bc}}$,
given in (\ref{operators}). These five operators have their own multiplicative
table; we quote below only some of the relevant products amongst them:%

\[
\mathsf{S}_{\mathsf{ab,ef}}^{(2a)}\mathsf{S}^{(2a)}{}^{\mathsf{ef}}%
{}_{,\mathsf{cd}}=-16\square\mathsf{P}_{\mathsf{ab,cd}}^{(2)},
\]

\[
\mathsf{R}_{\mathsf{ab,ef}}^{(1a)}\mathsf{R}^{(1a)}{}^{\mathsf{ef}}%
{}_{,\mathsf{cd}}=-4\square\mathsf{P}_{\mathsf{ab,cd}}^{(1m)},
\]

\[
\mathsf{P}_{\mathsf{ab,ef}}^{(2)}\mathsf{S}^{(2a)}{}^{\mathsf{ef}}%
{}_{,\mathsf{cd}}=\mathsf{S}_{\mathsf{ab,ef}}^{(2a)}\mathsf{P}^{(2)}%
{}^{\mathsf{ef}}{}_{,\mathsf{cd}}=\mathsf{S}_{\mathsf{ab,cd}}^{(2a)},
\]

\begin{equation}
\mathsf{P}_{\mathsf{ab,ef}}^{(1m)}\mathsf{R}^{(1a)}{}^{\mathsf{ef}}%
{}_{,\mathsf{cd}}=\mathsf{R}_{\mathsf{ab,ef}}^{(1a)}\mathsf{P}^{(1m)}%
{}^{\mathsf{ef}}{}_{,\mathsf{cd}}=\mathsf{R}_{\mathsf{ab,cd}}^{(1m)},
\end{equation}

\[
\mathsf{A}_{\mathsf{ae}}\mathsf{A}^{\mathsf{e}}{}_{\mathsf{b}}=-\square
\theta_{\mathsf{ab}},
\]

\[
\mathsf{B}_{\mathsf{a,ef}}\mathsf{B}_{\mathsf{c,}}{}^{\mathsf{ef}}%
=2\square(\theta_{\mathsf{ac}}+2\omega_{\mathsf{ac}}),
\]

\[
\mathsf{B}_{\mathsf{e,ab}}\mathsf{B}^{\mathsf{e}}{}_{,\mathsf{cd}}%
=2\square(\mathsf{P}_{\mathsf{ab,cd}}^{(1m)}+2\mathsf{P}_{\mathsf{ab,cd}%
}^{(0w)}),
\]

\[
\mathsf{D}_{\mathsf{a,ef}}\mathsf{D}_{\mathsf{c,}}{}^{\mathsf{ef}}%
=2\square^{2}\theta_{\mathsf{ac}}%
\]
and%

\[
\mathsf{D}_{\mathsf{e,ab}}\mathsf{D}^{\mathsf{e}}{}_{,\mathsf{cd}}%
=2\square^{2}\mathsf{P}_{\mathsf{ab,cd}}^{(1m)}.
\]

Here we have, just for the sake of calculational simplification, omitted the
$h^{\mathsf{c}}$\ component, dual of $\mathcal{H}_{\mathsf{ab}}%
=H_{[\mathsf{ab]}}$, in $M$, since we are not going to actually calculate the
propagators in this section, and we already can see the problem that is going
to appear with the reduced matrix. In the next section, where the propagators
for all field components will be derived, the anti-symmetric part will not be
left aside.

Thus, the wave operator acquires the form, without the anti-symmetric part of
$H_{\mathsf{ab}}$ as commented above:%

\begin{equation}
M=\left(
\begin{array}
[c]{ccc}%
M_{\mathsf{ab,cd}}^{yy} & M_{\mathsf{ab,c}}^{yy} & M_{\mathsf{ab,cd}}^{yh}\\
M_{\mathsf{a,cd}}^{yy} & M_{\mathsf{a,c}}^{yy} & M_{\mathsf{a,cd}}^{yh}\\
M_{\mathsf{ab,cd}}^{hy} & M_{\mathsf{ab,c}}^{hy} & M_{\mathsf{ab,cd}}^{hh}%
\end{array}
\right)  ,
\end{equation}
where%

\begin{align*}
M_{\mathsf{ab,cd}}^{yy}  &  =(2\mathsf{a}_{1}-2\mathsf{a}_{3}\square
)\mathsf{P}_{\mathsf{ab,cd}}^{(2)}+(2\mathsf{a}_{1}-\mathsf{a}_{3}\square
-2\xi\square)\mathsf{P}_{\mathsf{ab,cd}}^{(1m)}-(2\mathsf{a}_{1}%
+2\mathsf{a}_{3}\square)\mathsf{P}_{\mathsf{ab,cd}}^{(0s)}\\
&  -4\xi\square\mathsf{P}_{\mathsf{ab,cd}}^{(0w)}-2\sqrt{2}\mathsf{a}%
_{1}(\mathsf{P}_{\mathsf{ab,cd}}^{(0sw)}+\mathsf{P}_{\mathsf{ab,cd}}%
^{(0ws)})+\frac{\mathsf{a}_{4}}{2}(\mathsf{S}_{\mathsf{ab,cd}}^{(2a)}%
+\mathsf{R}_{\mathsf{ab,cd}}^{(1a)}),
\end{align*}

\[
M_{\mathsf{ab,c}}^{yy}=\mathsf{a}_{4}\mathsf{B}_{\mathsf{c,ab}}+(2\xi
-\mathsf{a}_{3})\mathsf{D}_{\mathsf{c,ab}},
\]

\[
M_{\mathsf{ab,cd}}^{yh}=\frac{k\square}{2}\mathsf{a}_{4}(\mathsf{P}%
_{\mathsf{ab,cd}}^{(2)}-\mathsf{P}_{\mathsf{ab,cd}}^{(0s)})+\frac{k}%
{4}\mathsf{a}_{1}(\mathsf{S}_{\mathsf{ab,cd}}^{(2a)}+\mathsf{R}%
_{\mathsf{ab,cd}}^{(1a)}),
\]

\[
M_{\mathsf{a,cd}}^{yy}=-\mathsf{a}_{4}\mathsf{B}_{\mathsf{a,bc}}%
+(2\xi-\mathsf{a}_{3})\mathsf{D}_{\mathsf{a,bc}},
\]

\begin{equation}
M_{\mathsf{a,c}}^{yy}=-(4\mathsf{a}_{1}+2\mathsf{a}_{3}\square+4\xi
\square)\theta_{\mathsf{a,c}}-(4\mathsf{a}_{1}+32\mathsf{a}_{2}\square
+12\mathsf{a}_{3}\square)\omega_{\mathsf{a,c}}+2\mathsf{a}_{4}\mathsf{A}%
_{\mathsf{a,c}},
\end{equation}

\[
M_{\mathsf{a,cd}}^{yh}=-\frac{k}{2}\mathsf{a}_{1}\mathsf{B}_{\mathsf{a,bc}%
}+k\mathsf{a}_{1}(\theta_{\mathsf{bc}}+\omega_{\mathsf{bc}})\partial
_{\mathsf{a}},
\]

\[
M_{\mathsf{ab,cd}}^{hy}=\frac{k\square}{2}\mathsf{a}_{4}(\mathsf{P}%
_{\mathsf{ab,cd}}^{(2)}-\mathsf{P}_{\mathsf{ab,cd}}^{(0s)})+\frac{k}%
{4}\mathsf{a}_{1}(\mathsf{S}_{\mathsf{ab,cd}}^{(2a)}+\mathsf{R}%
_{\mathsf{ab,cd}}^{(1a)}),
\]

\[
M_{\mathsf{ab,c}}^{hy}=\frac{k}{2}\mathsf{a}_{1}\mathsf{B}_{\mathsf{a,bc}%
}-k\mathsf{a}_{1}(\theta_{\mathsf{bc}}+\omega_{\mathsf{bc}})\partial
_{\mathsf{a}}%
\]
and%

\[
M_{\mathsf{ab,cd}}^{hh}=-\lambda\square\left(  \mathsf{P}_{\mathsf{ab,cd}%
}^{(1m)}+\mathsf{P}_{\mathsf{ab,cd}}^{(0s)}+\frac{1}{2}\mathsf{P}%
_{\mathsf{ab,cd}}^{(0w)}-\frac{\sqrt{2}}{2}(\mathsf{P}_{\mathsf{ab,cd}%
}^{(0sw)}+\mathsf{P}_{\mathsf{ab,cd}}^{(0ws)})\right)  .
\]

In order to write down the propagators of the model,%

\begin{equation}
\left\langle 0\right\vert T[\Phi(x)\Phi(y)]\left\vert 0\right\rangle =iM^{-1}%
\delta^{(3)}(x-y), \label{dezoito}%
\end{equation}
we need to calculate the inverse matrix, $M^{-1}$, of the wave operator. But,
here we face a problem: the matrix element $M_{\mathsf{ab,cd}}^{hh}$ does not have a
term in $\mathsf{P}_{\mathsf{ab,cd}}^{(2)}$, and we cannot find the inverse
element of this fundamental term (to compute the inverse, we need to close the
relation given in eq.(\ref{onze}), which not occur). 

At this point, a comment is worthy: the lack of invertibility of the wave operator, $M$, is
understandable and should be expected, once we are now adopting the
first-order formalism, where some of the gravity-field components are
non-dynamical, and so can rather be replaced in terms of the independent
components by means of the classical equations of motion, which acctually play
the role of constraints. This is a particularity of auxiliary fields appearing
in actions with local symmetry. This is indeed the case of gravity. We can see, in this manner, that a completely invertible theory, when
decomposed in terms of one gauge field and its torsion tensor components, has
difficulties when we adopt the version where the torsion is not taken as the
fundamental field, but rather work with the gauge field associated to Lorentz
local transformation that incorporates the torsion information (in a
Einstein-Cartan theory $\omega_{\mathsf{abc}}=\gamma_{\mathsf{abc}%
}-K_{\mathsf{abc}}$, where $\gamma_{\mathsf{abc}}$ is the "pure Riemannian",
without torsion, part and $K_{\mathsf{abc}}$\ is the contortion term). The
missing spin-2 term of the gravitational gauge field is incorporated into the
"Riemannian part" of the spin connection gauge field in the first-order formalism.

\section{Introducing the Torsion Terms}

In order to try to formulate a pure gauge model for planar gravitation, and yet
to understand the role torsion plays, we drop from the action (\ref{dez}) to a new action where the curvature terms are replaced by torsion:%

\begin{equation}
\mathcal{S=}\int d^{3}x\ e(\mathsf{a}_{1}\mathcal{R+}\mathsf{a}_{2}%
\mathcal{T}_{\alpha\beta\gamma}\mathcal{T}^{\alpha\beta\gamma}+\mathsf{a}%
_{3}\mathcal{T}_{\alpha\beta\gamma}\mathcal{T}^{\beta\gamma\alpha}%
+\mathsf{a}_{4}\mathcal{T}_{\alpha\beta}{}^{\beta}\mathcal{T}^{\alpha}%
{}_{\gamma}{}^{\gamma}+\mathsf{a}_{5}\mathcal{L}_{CS}). \label{treze}%
\end{equation}

 $\mathcal{L}_{CS}$\ being the usual Chern-Simons term, given in eq. (\ref{quatorze}).
$\mathsf{a}_{1}$, $\mathsf{a}_{2}$, $\mathsf{a}_{3}$\ and $\mathsf{a}_{4}$ are
free coefficients, whereas $\mathsf{a}_{5}$ is the Chern-Simons parameter. See
reference \cite{Sezgin:1979zf} for a complete discussion about the specific torsion terms. From now on, all
our calculations and results refer to the action (\ref{treze}). In our final
section, we shall make a comment on the possibility of introducing a term which is
linear in the torsion \cite{Zanelli:2002qm}.

We consider equations (\ref{nove}), (\ref{doze}) and (\ref{sete}), and the
decompositions (\ref{quinze}), (\ref{dezesseis}) and (\ref{dezessete}), with
the weak expansion (\ref{vintedois}) and equations (\ref{vintetres}) and
(\ref{vintequatro}), we can rewrite the action (\ref{treze}), introducing the
gauge-fixing term,%

\begin{equation}
\mathcal{L}_{GF-diff}=\lambda F_{\mathsf{a}}F^{\mathsf{a}}\ ,\ F_{\mathsf{a}%
}=k\partial^{\mathsf{b}}\left(  H_{\mathsf{ba}}-\frac{1}{2}\eta_{\mathsf{ba}%
}H_{\mathsf{c}}{}^{\mathsf{c}}\right)  ,
\end{equation}

in the linearized form below:%

\begin{equation}
\mathcal{S}\mathcal{=}\int d^{3}x\frac{1}{2}\Phi^{T}M\Phi\ ,\ \Phi=\left(
\begin{array}
[c]{c}%
h^{\mathsf{cd}}\\
h^{\mathsf{c}}\\
y^{\mathsf{cd}}\\
y^{\mathsf{c}}%
\end{array}
\right)  .
\end{equation}

As before, we express the wave operator, $M$, in terms of the extended
spin-projection operators. In addition to the operators listed above,
there appear two new operators:%

\begin{equation}
\theta_{\mathsf{ab}}\partial_{\mathsf{c}}\text{ \ \ \ \ \ and \ \ \ \ }%
\omega_{\mathsf{ab}}\partial_{\mathsf{c}},
\end{equation}
which, together with the old ones, completely close the algebra.

This yields the form below for the wave operator:%

\begin{equation}
M=\left(
\begin{array}
[c]{cccc}%
M_{\mathsf{ab,cd}}^{hh} & M_{\mathsf{ab,c}}^{hh} & M_{\mathsf{ab,cd}}^{hy} &
M_{\mathsf{ab,c}}^{hy}\\
M_{\mathsf{a,cd}}^{hh} & M_{\mathsf{a,c}}^{hh} & M_{\mathsf{a,cd}}^{hy} &
M_{\mathsf{a,c}}^{hy}\\
M_{\mathsf{ab,cd}}^{yh} & M_{\mathsf{ab,c}}^{yh} & M_{\mathsf{ab,cd}}^{yy} &
M_{\mathsf{ab,c}}^{yy}\\
M_{\mathsf{a,cd}}^{yh} & M_{\mathsf{a,c}}^{yh} & M_{\mathsf{a,cd}}^{yy} &
M_{\mathsf{a,c}}^{yy}%
\end{array}
\right)  ,
\end{equation}
where%

\begin{eqnarray}
M_{\mathsf{ab,cd}}^{hh}  &=&\frac{k^{2}}{2}\square(\mathsf{a}_{3}%
-2\mathsf{a}_{2})\mathsf{P}_{\mathsf{ab,cd}}^{(2)}+\frac{k^{2}}{4}%
\square(\mathsf{a}_{3}-2\mathsf{a}_{2}-\mathsf{a}_{4}-4\lambda)\mathsf{P}%
_{\mathsf{ab,cd}}^{(1m)}\nonumber\\
&& +\frac{k^{2}}{2}\square(\mathsf{a}_{3}-2\mathsf{a}_{2}-2\mathsf{a}%
_{4}-2\lambda)\mathsf{P}_{\mathsf{ab,cd}}^{(0s)}-(\frac{k^{2}}{2}%
\square\lambda)\mathsf{P}_{\mathsf{ab,cd}}^{(0w)}\nonumber\\
&&  -(\frac{\sqrt{2}}{2}k^{2}\square\lambda)(\mathsf{P}_{\mathsf{ab,cd}%
}^{(0sw)}+\mathsf{P}_{\mathsf{ab,cd}}^{(0ws)})-\frac{\mathsf{k}^{2}}%
{2}\mathsf{a}_{5}(\mathsf{S}_{\mathsf{ab,cd}}^{(2a)}+\mathsf{R}%
_{\mathsf{ab,cd}}^{(1a)}),
\end{eqnarray}

\begin{equation}
M_{\mathsf{ab,c}}^{hh}=-(\frac{k^{2}}{2}\mathsf{a}_{5})\mathsf{B}%
_{\mathsf{c,ab}}+\frac{k^{2}}{4}(\mathsf{a}_{3}-2\mathsf{a}_{2}-\mathsf{a}%
_{4}+4\lambda)\mathsf{D}_{\mathsf{c,ab}},
\end{equation}

\begin{eqnarray}
M_{\mathsf{ab,cd}}^{hy}  &=&\frac{k}{2}(\square\mathsf{a}_{6}-2\mathsf{a}%
_{5})\mathsf{P}_{\mathsf{ab,cd}}^{(2)}-(k\mathsf{a}_{5})\mathsf{P}%
_{\mathsf{ab,cd}}^{(1m)}-\frac{k}{2}(\square\mathsf{a}_{6}+2\mathsf{a}%
_{5})\mathsf{P}_{\mathsf{ab,cd}}^{(0s)}\nonumber\\
&&  -(k\mathsf{a}_{5})\mathsf{P}_{\mathsf{ab,cd}}^{(0w)}+\frac{k}{4}%
(\mathsf{a}_{1}+2\mathsf{a}_{2}-2\mathsf{a}_{3})(\mathsf{S}_{\mathsf{ab,cd}%
}^{(2a)}+\mathsf{R}_{\mathsf{ab,cd}}^{(1a)}),
\end{eqnarray}

\begin{equation}
M_{\mathsf{ab,c}}^{hy}=\frac{k}{2}(\mathsf{a}_{1}-2\mathsf{a}_{2}%
-2\mathsf{a}_{4})\mathsf{B}_{\mathsf{c,ab}}-k(\mathsf{a}_{1}-2\mathsf{a}%
_{2}-2\mathsf{a}_{4})(\theta_{\mathsf{ab}}+\omega_{\mathsf{ab}})\partial
_{\mathsf{c}},
\end{equation}

\begin{equation}
M_{\mathsf{a,cd}}^{hh}=(\frac{k^{2}}{2}\mathsf{a}_{5})\mathsf{B}%
_{\mathsf{a,cd}}+\frac{k^{2}}{4}(\mathsf{a}_{3}-2\mathsf{a}_{2}-\mathsf{a}%
_{4}+4\lambda)\mathsf{D}_{\mathsf{a,cd}},
\end{equation}

\begin{equation}
M_{\mathsf{a,c}}^{hh}=\frac{k^{2}}{2}\square(\mathsf{a}_{3}-2\mathsf{a}%
_{2}-\mathsf{a}_{4}-4\lambda)\theta_{\mathsf{a,c}}-(k^{2}\square
)(2\mathsf{a}_{2}+\mathsf{a}_{3})\omega_{\mathsf{a,c}}-(k^{2}\mathsf{a}%
_{5})\mathsf{A}_{\mathsf{a,c}},
\end{equation}

\begin{equation}
M_{\mathsf{a,cd}}^{hy}=-\frac{k}{2}(\mathsf{a}_{1}+2\mathsf{a}_{2}%
)\mathsf{B}_{\mathsf{a,bc}}+k(\mathsf{a}_{1}-2\mathsf{a}_{2}-2\mathsf{a}%
_{3})(\theta_{\mathsf{bc}}+\omega_{\mathsf{bc}})\partial_{\mathsf{a}},
\end{equation}

\begin{equation}
M_{\mathsf{a,c}}^{hy}=(2k\mathsf{a}_{5})\theta_{\mathsf{a,c}}+k(2\mathsf{a}%
_{5}-\square\mathsf{a}_{6})\omega_{\mathsf{a,c}}+k(\mathsf{a}_{1}%
-2\mathsf{a}_{2}-2\mathsf{a}_{4})\mathsf{A}_{\mathsf{a,c}},
\end{equation}

\begin{eqnarray}
M_{\mathsf{ab,cd}}^{yh}  &=&\frac{k}{2}(\square\mathsf{a}_{6}-2\mathsf{a}%
_{5})\mathsf{P}_{\mathsf{ab,cd}}^{(2)}-(k\mathsf{a}_{5})\mathsf{P}%
_{\mathsf{ab,cd}}^{(1m)}-\frac{k}{2}(\square\mathsf{a}_{6}+2\mathsf{a}%
_{5})\mathsf{P}_{\mathsf{ab,cd}}^{(0s)}\nonumber\\
&&  -(k\mathsf{a}_{5})\mathsf{P}_{\mathsf{ab,cd}}^{(0w)}+\frac{k}{4}%
(\mathsf{a}_{1}+2\mathsf{a}_{2}-2\mathsf{a}_{3})(\mathsf{S}_{\mathsf{ab,cd}%
}^{(2a)}+\mathsf{R}_{\mathsf{ab,cd}}^{(1a)}),
\end{eqnarray}

\begin{equation}
M_{\mathsf{ab,c}}^{yh}=\frac{k}{2}(\mathsf{a}_{1}+2\mathsf{a}_{2}%
)\mathsf{B}_{\mathsf{c,ab}}-k(\mathsf{a}_{1}-2\mathsf{a}_{2}-2\mathsf{a}%
_{3})(\theta_{\mathsf{ab}}+\omega_{\mathsf{ab}})\partial_{\mathsf{c}},
\end{equation}

\begin{eqnarray}
M_{\mathsf{ab,cd}}^{yy}  &=&2(\mathsf{a}_{1}+2\mathsf{a}_{2}-\mathsf{a}%
_{3})\mathsf{P}_{\mathsf{ab,cd}}^{(2)}+2(\mathsf{a}_{1}+2\mathsf{a}%
_{2}-\mathsf{a}_{3})\mathsf{P}_{\mathsf{ab,cd}}^{(1m)}\nonumber\\
&&  +2(6\mathsf{a}_{2}+5\mathsf{a}_{3}-\mathsf{a}_{1})\mathsf{P}%
_{\mathsf{ab,cd}}^{(0s)}+4(2\mathsf{a}_{2}+\mathsf{a}_{3})\mathsf{P}%
_{\mathsf{ab,cd}}^{(0w)}\nonumber\\
&&  +2\sqrt{2}(2\mathsf{a}_{2}+3\mathsf{a}_{3}-\mathsf{a}_{1})(\mathsf{P}%
_{\mathsf{ab,cd}}^{(0sw)}+\mathsf{P}_{\mathsf{ab,cd}}^{(0ws)})+(\frac
{\mathsf{a}_{6}}{2})(\mathsf{S}_{\mathsf{ab,cd}}^{(2a)}+\mathsf{R}%
_{\mathsf{ab,cd}}^{(1a)}),\nonumber\\
\end{eqnarray}

\begin{equation}
M_{\mathsf{ab,c}}^{yy}=\mathsf{a}_{6}\mathsf{B}_{\mathsf{c,ab}},
\end{equation}

\begin{equation}
M_{\mathsf{a,cd}}^{yh}=-\frac{k}{2}(\mathsf{a}_{1}-2\mathsf{a}_{2}%
-2\mathsf{a}_{4})\mathsf{B}_{\mathsf{a,bc}}+k(\mathsf{a}_{1}-2\mathsf{a}%
_{2}-2\mathsf{a}_{4})(\theta_{\mathsf{bc}}+\omega_{\mathsf{bc}})\partial
_{\mathsf{a}},
\end{equation}

\begin{equation}
M_{\mathsf{a,c}}^{yh}=(2k\mathsf{a}_{5})\theta_{\mathsf{a,c}}+k(2\mathsf{a}%
_{5}-\square\mathsf{a}_{6})\omega_{\mathsf{a,c}}+k(\mathsf{a}_{1}%
-2\mathsf{a}_{2}-2\mathsf{a}_{4})\mathsf{A}_{\mathsf{a,c}},
\end{equation}

\begin{equation}
M_{\mathsf{a,cd}}^{yy}=-\mathsf{a}_{6}\mathsf{B}_{\mathsf{a,cd}}%
\end{equation}
and%

\begin{eqnarray}
M_{\mathsf{a,c}}^{yy}  &=& 4(2\mathsf{a}_{2}+2\mathsf{a}_{4}-\mathsf{a}%
_{1}-\mathsf{a}_{3})\theta_{\mathsf{a,c}}+4(2\mathsf{a}_{2}+2\mathsf{a}%
_{4}-\mathsf{a}_{1}-\mathsf{a}_{3})\omega_{\mathsf{a,c}}\nonumber\\
&&  +(2\mathsf{a}_{6})\mathsf{A}_{\mathsf{a,c}}.
\end{eqnarray}

Once all operators have been identified, we finally come to the
task of computing the inverses. This is what we shall do next.

\section{Propagators and Excitation Modes}

In order to calculate the propagators, eq. (\ref{dezoito}), we use a
straightforward, but lengthy, procedure in terms of which we decompose the
matrix $M$\ into four sectors, namely:%

\begin{equation}
M=\left(
\begin{array}
[c]{cc}%
M^{hh} & M^{hy}\\
M^{yh} & M^{yy}%
\end{array}
\right)  .
\end{equation}

Thus the inverse matrix $M^{-1}$\ can be written as:%

\begin{equation}
M^{-1}=\left(
\begin{array}
[c]{cc}%
M^{HH} & M^{HY}\\
M^{YH} & M^{YY}%
\end{array}
\right)  ,
\end{equation}

where its blocks are given by:%

\begin{align}
M^{HH}  &  =[M^{hh}-M^{hy}(M^{yy})^{-1}M^{yh}]^{-1}.\nonumber\\
M^{HY}  &  =-(M^{hh})^{-1}M^{hy}M^{YY}.\label{dezenove}\\
M^{YH}  &  =-(M^{yy})^{-1}M^{yh}M^{HH}.\nonumber\\
M^{YY}  &  =[M^{yy}-M^{yh}(M^{hh})^{-1}M^{hy}]^{-1}.\nonumber
\end{align}

Once the propagators are read off, we must check the tree-level unitarity of
the theory. To this, we have to analyse the residues of the current-current
transition amplitude in momentum space, given by the saturated propagator
after a Fourier transformation. The sources that saturate the propagators can
be expanded in terms of a complete basis in the momentum space as follows:%

\begin{align}
S_{\mu\nu}  &  =c%
\acute{}%
_{1}p_{\mu}p_{\nu}+c%
\acute{}%
_{2}p_{\mu}q_{\nu}+c%
\acute{}%
_{3}p_{\mu}\varepsilon_{\nu}+c%
\acute{}%
_{4}q_{\mu}p_{\nu}+c%
\acute{}%
_{5}q_{\mu}q_{\nu}\\
&  +c%
\acute{}%
_{6}q_{\mu}\varepsilon_{\nu}+c%
\acute{}%
_{7}\varepsilon_{\mu}p_{\nu}+c%
\acute{}%
_{8}\varepsilon_{\mu}q_{\nu}+c%
\acute{}%
_{9}\varepsilon_{\mu}\varepsilon_{\nu},\nonumber
\end{align}
where $p_{\mu}=(p_{0},-\overrightarrow{p})$, $q_{\mu}=(p_{0},\overrightarrow
{p})$\ and $\varepsilon_{\mu}=(0,-\overrightarrow{\varepsilon})$\ are linearly
independent vectors that satisfy the conditions:%

\begin{align}
p_{\mu}p^{\mu}  &  =q_{\mu}q^{\mu}=m^{2}.\nonumber\\
p_{\mu}q^{\mu}  &  =p_{0}^{2}+\overrightarrow{p}^{2}\neq0.\\
p_{\mu}\varepsilon^{\mu}  &  =q_{\mu}\varepsilon^{\mu}=0.\nonumber\\
\varepsilon_{\mu}\varepsilon^{\mu}  &  =-1.\nonumber
\end{align}

These conditions and the symmetry requirements of the theory split the
sources, $S_{\mu\nu}$, in a symmetric and an antisymmetric part:%

\begin{align}
S_{S\mu\nu}  &  =S_{(\mu\nu)}=c_{1}p_{\mu}p_{\nu}+c_{2}(p_{\mu}q_{\nu}+q_{\mu
}p_{\nu})+c_{3}(p_{\mu}\varepsilon_{\nu}+\varepsilon_{\mu}p_{\nu})\\
&  +c_{4}q_{\mu}q_{\nu}+c_{5}(q_{\mu}\varepsilon_{\nu}+\varepsilon_{\mu}%
q_{\nu})+c_{6}\varepsilon_{\mu}\varepsilon_{\nu}\nonumber
\end{align}
and%

\begin{align}
A_{S\mu\nu}  &  =S_{[\mu\nu]}=d_{1}(p_{\mu}q_{\nu}-q_{\mu}p_{\nu}%
)+d_{2}(p_{\mu}\varepsilon_{\nu}-\varepsilon_{\mu}p_{\nu})\\
&  +d_{3}(q_{\mu}\varepsilon_{\nu}-\varepsilon_{\mu}q_{\nu}),\nonumber
\end{align}
where $c_{1}=c%
\acute{}%
_{1},$ $c_{2}=\frac{c%
\acute{}%
_{2}+c%
\acute{}%
_{4}}{2},$ $c_{3}=\frac{c%
\acute{}%
_{3}+c%
\acute{}%
_{7}}{2},$ $c_{4}=c%
\acute{}%
_{5},$ $c_{5}=\frac{c%
\acute{}%
_{6}+c%
\acute{}%
_{8}}{2},$ $c_{6}=c%
\acute{}%
_{9}$ $d_{1}=\frac{c%
\acute{}%
_{2}-c%
\acute{}%
_{4}}{2},$ $d_{2}=\frac{c%
\acute{}%
_{3}-c%
\acute{}%
_{7}}{2},$ and $d_{3}=\frac{c%
\acute{}%
_{6}-c%
\acute{}%
_{8}}{2}$.

The currente-current transition amplitude is written as:%

\begin{align}
\mathcal{A}  &  =\left(
\begin{array}
[c]{cc}%
\tau^{\ast} & \rho^{\ast}%
\end{array}
\right)  \left(
\begin{array}
[c]{cc}%
M^{HH} & M^{HY}\\
M^{YH} & M^{YY}%
\end{array}
\right)  \left(
\begin{array}
[c]{c}%
\tau\\
\rho
\end{array}
\right)  \Rightarrow\\
\mathcal{A}  &  =\tau^{\ast}M^{HH}\tau+\tau^{\ast}M^{HY}\rho+\rho^{\ast}%
M^{YH}\tau+\rho^{\ast}M^{YY}\rho,\nonumber
\end{align}
where $\tau$\ is the source to the $h$\ fields and $\rho$\ the source to the
$y$\ fields.

$\mathcal{A}$ can then be cast into the form below:%

\begin{align}
\mathcal{A}  &  =t^{\mathsf{ab}^{\ast}}M_{\mathsf{ab,cd}}^{HH}t^{\mathsf{cd}%
}+t^{\mathsf{ab}^{\ast}}M_{\mathsf{ab,c}}^{HH}t^{\mathsf{c}}+t^{\mathsf{a}%
^{\ast}}M_{\mathsf{a,cd}}^{HH}t^{\mathsf{cd}}+t^{\mathsf{a}^{\ast}%
}M_{\mathsf{a,c}}^{HH}t^{\mathsf{c}}\nonumber\\
&  +t^{\mathsf{ab}^{\ast}}M_{\mathsf{ab,cd}}^{HY}r^{\mathsf{cd}}%
+t^{\mathsf{ab}^{\ast}}M_{\mathsf{ab,c}}^{HY}r^{\mathsf{c}}+t^{\mathsf{a}%
^{\ast}}M_{\mathsf{a,cd}}^{HY}r^{\mathsf{cd}}+t^{\mathsf{a}^{\ast}%
}M_{\mathsf{a,c}}^{HY}r^{\mathsf{c}}\\
&  +r^{\mathsf{ab}^{\ast}}M_{\mathsf{ab,cd}}^{YH}t^{\mathsf{cd}}%
+r^{\mathsf{ab}^{\ast}}M_{\mathsf{ab,c}}^{YH}t^{\mathsf{c}}+r^{\mathsf{a}%
^{\ast}}M_{\mathsf{a,cd}}^{YH}t^{\mathsf{cd}}+r^{\mathsf{a}^{\ast}%
}M_{\mathsf{a,c}}^{YH}t^{\mathsf{c}}\nonumber\\
&  +r^{\mathsf{ab}^{\ast}}M_{\mathsf{ab,cd}}^{YY}r^{\mathsf{cd}}%
+r^{\mathsf{ab}^{\ast}}M_{\mathsf{ab,c}}^{YY}r^{\mathsf{c}}+r^{\mathsf{a}%
^{\ast}}M_{\mathsf{a,cd}}^{YY}r^{\mathsf{cd}}+r^{\mathsf{a}^{\ast}%
}M_{\mathsf{a,c}}^{YY}r^{\mathsf{c}},\nonumber
\end{align}
where $t^{\mathsf{cd}}=\tau^{(\mathsf{cd)}}$,\ $t^{\mathsf{c}}=\frac{1}%
{2}\epsilon^{\mathsf{cde}}T_{\mathsf{de}}$ with $T_{\mathsf{de}}=\tau
_{\lbrack\mathsf{de}]}$ and $r^{\mathsf{cd}}=\rho^{(\mathsf{cd)}}%
$,\ $r^{\mathsf{c}}=\frac{1}{2}\epsilon^{\mathsf{cde}}R_{\mathsf{de}}$ with
$R_{\mathsf{de}}=\rho_{\lbrack\mathsf{de}]}$.

Due to the source constraints, $p_{\mathsf{c}}t^{\mathsf{cd}}=0$,
$p_{\mathsf{c}}T^{\mathsf{cd}}=0$, $p_{\mathsf{c}}r^{\mathsf{cd}}=0$ and
$p_{\mathsf{c}}R^{\mathsf{cd}}=0$, only the projectors $\mathsf{P}%
_{\mathsf{ab,cd}}^{(2)}$, $\mathsf{P}_{\mathsf{ab,cd}}^{(0s)}$, $\mathsf{S}%
_{\mathsf{ab,cd}}^{(2a)}$, $\theta_{\mathsf{ab}}\partial_{\mathsf{c}}$and
$\omega_{\mathsf{a,c}}$,give non-vanishing contributions to the amplitude.

For a massless pole, or for a massive pole in the rest frame (where $p_{\mu
}=(m,0)$, $q_{\mu}=(m,0)$\ and $\varepsilon_{\mu}=(0,-\overrightarrow
{\varepsilon})$), only the projectors $\mathsf{P}_{\mathsf{ab,cd}}^{(2)}$ and
$\mathsf{P}_{\mathsf{ab,cd}}^{(0s)}$\ survive and contribute.

With the restrictions above, the amplitude reads:%

\begin{align}
\mathcal{A}  &  =<H2H2_{(2)}>t^{\mathsf{ab}^{\ast}}\mathsf{P}_{\mathsf{ab,cd}%
}^{(2)}t^{\mathsf{cd}}+<H2H2_{(0s)}>t^{\mathsf{ab}^{\ast}}\mathsf{P}%
_{\mathsf{ab,cd}}^{(0s)}t^{\mathsf{cd}}\nonumber\\
+  &  <H2Y2_{(2)}>t^{\mathsf{ab}^{\ast}}\mathsf{P}_{\mathsf{ab,cd}}%
^{(2)}r^{\mathsf{cd}}+<H2Y2_{(0s)}>t^{\mathsf{ab}^{\ast}}\mathsf{P}%
_{\mathsf{ab,cd}}^{(0s)}r^{\mathsf{cd}}\\
+  &  <Y2H2_{(2)}>r^{\mathsf{ab}^{\ast}}\mathsf{P}_{\mathsf{ab,cd}}%
^{(2)}t^{\mathsf{cd}}+<Y2H2_{(0s)}>r^{\mathsf{ab}^{\ast}}\mathsf{P}%
_{\mathsf{ab,cd}}^{(0s)}t^{\mathsf{cd}}\nonumber\\
+  &  <Y2Y2_{(2)}>r^{\mathsf{ab}^{\ast}}\mathsf{P}_{\mathsf{ab,cd}}%
^{(2)}r^{\mathsf{cd}}+<Y2Y2_{(0s)}>r^{\mathsf{ab}^{\ast}}\mathsf{P}%
_{\mathsf{ab,cd}}^{(0s)}r^{\mathsf{cd}},\nonumber
\end{align}
where $<H2H2_{(2)}>$ is the symmetric rank-2 ($H2$ in $H2H2_{(2)}$)
gravitational field propagator associated to the operator $\mathsf{P}%
_{\mathsf{ab,cd}}^{(2)}$ ($_{(2)}$ in $H2H2_{(2)}$). The other coefficients
have analogous meaning. Explicitly writing the sources, we get:%

\begin{align}
\mathcal{A}  &  =\frac{1}{2}(<H2H2_{(2)}>+<H2H2_{(0s)}>)\left\vert
c_{6t}\right\vert ^{2}\nonumber\\
+\frac{1}{2}(  &  <H2Y2_{(2)}>+<H2Y2_{(0s)}>)c_{6t}^{\ast}c_{6r}%
\label{vinte}\\
+\frac{1}{2}(  &  <Y2H2_{(2)}>+<Y2H2_{(0s)}>)c_{6r}^{\ast}c_{6t}\nonumber\\
+\frac{1}{2}(  &  <Y2Y2_{(2)}>+<Y2Y2_{(0s)}>)\left\vert c_{6r}\right\vert
^{2}\nonumber
\end{align}
where $t$ and $r$ in the $c$ mean the source associated to the particular term.

We must now replace the results obtained by the procedure described in
(\ref{dezenove}) into (\ref{vinte}). Before, explicitly we put our results,
the following comments should be done:

\begin{enumerate}
\item With the whole set of action parameters, $\mathsf{a}_{1}$,
$\mathsf{a}_{2}$, $\mathsf{a}_{3}$, $\mathsf{a}_{4}$ $\mathsf{a}_{5}$ and
$\lambda$, different from zero, our computational algebraic facilities failed
in attaining an analytical result, due to the extension of the resulting expressions.

\item Considering the Chern-Simons term, $\mathsf{a}_{5}$,we obtained the
following behaviour in the denominator of the propagator:

\begin{itemize}
\item With $\mathsf{a}_{1}=0$, we have terms proportional to $p^{22}$.

\item The lowest power, $p^{6}$, occurs with $\mathsf{a}_{1}=\mathsf{a}%
_{2}=\mathsf{a}_{4}=0$, only $\mathsf{a}_{3}$\ and $\mathsf{a}_{5}$\ are considered.

\item With $\mathsf{a}_{3}=0$, we do not have\ an invertible case.
\end{itemize}

\item Without the Chern-Simons term, $\mathsf{a}_{5}=0$, we obtain, in all
invertible cases, a power $p^{2}$. This is not a straightforward result; we
may justify it by pointing out that Chern-Simons contributes a term quadratic
in the spin connection with a space-time derivative, whereas the scalar
curvature contributes a term that mixes $H$ with $\omega$. Setting
$\mathsf{a}_{5}$\ to zero, we suppres $\omega-\omega$\ terms with a
derivative, and so we unavoidably reduce the powers in the momentum appearing
in the propagators.
\end{enumerate}

We then consider in (\ref{vinte}) only the cases with $\mathsf{a}_{5}=0$.

Anyway, we have a procedure that works out for all possibilities in parameter
space (once we keep $\mathsf{a}_{6}=0$; we come back to this point in our
Concluding Comments). We simply report here the cases with $\mathsf{a}_{3}$,
and $\mathsf{a}_{1}$\ and $\mathsf{a}_{3}$, different from zero to have an
illustration of how our general procedure works.

The least invertible case occurs by considering only $\mathsf{a}_{3}$
different from zero in the action. In this case, the relevant propagators read:%

\begin{align}
H2H2_{(2)}  &  =\frac{2}{3k^{2}p^{2}\mathsf{a}_{3}}i.\nonumber\\
H2H2_{(0s)}  &  =-\frac{2}{k^{2}p^{2}\mathsf{a}_{3}}i.\nonumber\\
H2Y2_{(2)}  &  =H2Y2_{(0s)}=Y2H2_{(2)}=Y2H2_{(0s)}=0.\\
Y2Y2_{(2)}  &  =\frac{1}{6\mathsf{a}_{3}}i.\nonumber\\
Y2Y2_{(0s)}  &  =0\nonumber
\end{align}
and the saturated amplitude is as given below,%

\begin{equation}
\mathcal{A}=\left(  -\frac{2}{3k^{2}p^{2}\mathsf{a}_{3}}\left\vert
c_{6}\right\vert _{tt}^{2}+\frac{1}{12\mathsf{a}_{3}}\left\vert c_{6}%
\right\vert _{rr}^{2}\right)  i.
\end{equation}

We notice in this expression that the massless pole comes from the $h$-block
and has contributions from the spin-0 and the spin-2 sectors.

Then, by calculating the imaginary part of the residue of the amplitude at the
massless pole, we get:%

\begin{equation}
\operatorname{Im}(res\mathcal{A})=\operatorname{Im}\left(  \lim_{p^{2}%
\rightarrow0}[p^{2}\mathcal{A}]\right)  =-\frac{2\left\vert c_{6}\right\vert
_{tt}^{2}}{3k^{2}\mathsf{a}_{3}}.
\end{equation}

From the requirement of having positive-definite residue at the pole, we must
have $\mathsf{a}_{3}<0$.

Considering now the addition of the scalar of curvature term $\mathsf{a}_{1}$,
we get:%

\begin{align}
H2H2_{(2)}  &  =\frac{2(\mathsf{a}_{3}-\mathsf{a}_{1})}{k^{2}p^{2}%
\mathsf{(}3\mathsf{a}_{3}^{2}+\mathsf{a}_{1}^{2}-3\mathsf{a}_{3}\mathsf{a}%
_{1})}i\nonumber\\
H2H2_{(0s)}  &  =-\frac{2(\mathsf{a}_{3}+\mathsf{a}_{1})}{k^{2}p^{2}%
\mathsf{(\mathsf{a}_{3}^{2}-\mathsf{a}_{1}^{2}+\mathsf{a}_{3}\mathsf{a}_{1})}%
}i\nonumber\\
H2Y2_{(2)}  &  =H2Y2_{(0s)}=Y2H2_{(2)}=Y2H2_{(0s)}=0\\
Y2Y2_{(2)}  &  =\frac{\mathsf{a}_{3}}{2(3\mathsf{a}_{3}^{2}+\mathsf{\mathsf{a}%
_{1}^{2}}-3\mathsf{\mathsf{a}_{3}\mathsf{a}_{1}})}i\nonumber\\
Y2Y2_{(0s)}  &  =0\nonumber
\end{align}
and the amplitude becomes:%

\begin{equation}
\mathcal{A}=\left(  -\frac{2}{k^{2}p^{2}}\times\frac{\mathsf{\mathsf{a}%
_{3}^{3}}}{3\mathsf{a}_{3}^{4}-5\mathsf{\mathsf{a}_{3}^{2}\mathsf{a}_{1}%
^{2}+4\mathsf{a}_{3}a}_{1}^{3}-\mathsf{a}_{1}^{4}}\left\vert c_{6}\right\vert
_{tt}^{2}+\frac{\mathsf{a}_{3}}{2(3\mathsf{a}_{3}^{2}+\mathsf{\mathsf{a}%
_{1}^{2}}-3\mathsf{\mathsf{a}_{3}\mathsf{a}_{1}})}\left\vert c_{6}\right\vert
_{rr}^{2}\right)  i.\nonumber
\end{equation}

We can see that the structure of the amplitude is not changed, with the pole
having contributions from the same spin sectors. The parameters relations now reads:%

\begin{equation}
\operatorname{Im}(res\mathcal{A})=\operatorname{Im}\left(  \lim_{p^{2}%
\rightarrow0}[p^{2}\mathcal{A}]\right)  =-\frac{2}{k^{2}}\frac
{\mathsf{\mathsf{a}_{3}^{3}}}{3\mathsf{a}_{3}^{4}-5\mathsf{\mathsf{a}_{3}%
^{2}\mathsf{a}_{1}^{2}+4\mathsf{a}_{3}a}_{1}^{3}-\mathsf{a}_{1}^{4}}\left\vert
c_{6}\right\vert _{tt}^{2}. \label{vinteum}%
\end{equation}

The denominator in (\ref{vinteum}) can be written as:%

\begin{equation}
\mathsf{(\mathsf{a}_{3}^{2}+\mathsf{a}_{3}\mathsf{a}_{1}-\mathsf{a}_{1}%
^{2})(3\mathsf{a}_{3}^{2}-3\mathsf{a}_{3}\mathsf{a}_{1}+\mathsf{a}_{1}^{2}).}%
\end{equation}

The binomial $\mathsf{3\mathsf{a}_{3}^{2}-3\mathsf{a}_{3}\mathsf{a}%
_{1}+\mathsf{a}_{1}^{2}}$ has complex roots and is greater than zero.

The requirement of having positive-definite residue at the pole implies (with
$\mathsf{a}_{3}<0$) $\mathsf{\mathsf{a}_{1}^{2}-\mathsf{a}_{3}\mathsf{a}%
_{1}-\mathsf{a}_{3}^{2}<0}$. And the scalar term must obey $\frac{1+\sqrt{5}%
}{2}\mathsf{\mathsf{a}_{3}\approx1.618\mathsf{a}_{3}}<\mathsf{\mathsf{a}_{1}%
<}\frac{1-\sqrt{5}}{2}\mathsf{\mathsf{a}_{3}\approx-0.618\mathsf{a}_{3}}$.

The case where all parameters (with exception to $\mathsf{a}_{5}$) are
different from zero brings only new algebraic corrections to the amplitude,
without changing its structure. The relations among the parameters become very
cumbersome, due to the considerable number of parameters involved, so that
many hypotheses must be done.

\section{Concluding Comments}

In the course of the calculations we report in this work, if we complete the
action (\ref{treze}) by adjoining the term $\mathsf{a}_{6}\epsilon^{\mu
\nu\lambda}\mathcal{T}_{\mu\nu}{}^{\mathsf{\mathsf{a}}}e_{\lambda}%
{}^{\mathsf{b}}\eta_{\mathsf{\mathsf{ab}}}=\mathsf{a}_{6}\epsilon^{\mu
\nu\lambda}\mathcal{T}_{\mu\nu}{}^{\mathsf{\alpha}}e_{\alpha}{}^{\mathsf{a}%
}e_{\lambda}{}^{\mathsf{b}}\eta_{\mathsf{\mathsf{ab}}}=\mathsf{a}_{6}%
\epsilon^{\mu\nu\lambda}\mathcal{T}_{\mu\nu\lambda}$ \cite{Zanelli:2002qm},
a\ problem shows up: though our procedure of introducing the spin operators
works, the propagators could not be found in their generality (with all the
six coefficients $\mathsf{a}_{i}$) even with the help of algebraic computation
techniques. However, we found out that, once any of the $\mathsf{a}_{i}$ are
set to zero, we succeed in reading off the propagators, even if they display
higher powers in the momentum. It is worthwhile to mention here that this
linear term in the torsion combines with the Chern-Simons action to give a
rich structure of poles in the propagators. We do not report these results
here because this investigation is the matter of a forthcoming publication
\cite{futurohelay}. The situation gets better when we discovered that, ruling
out the Chern-Simons term, we get only simple poles in the terms that
contribute to the amplitude. Very surprising was the discovery of the very
different role the torsion terms ($\mathsf{a}_{2}$ and $\mathsf{a}_{3}$) play,
being $\mathsf{a}_{3}$ fundamental to compute the inverse matrix, which is not
the case for $\mathsf{a}_{2}$.

We see that the physical poles are all massless. It is worthy to note that, in
\cite{Boldo:1999qw}, we get only physical mass poles. The unitarity condition
for the physical poles demand that $\mathsf{a}_{3}<0$ and this implies in that
the parameter that governs the scalar curvature must obey the condition
$\frac{1+\sqrt{5}}{2}\mathsf{\mathsf{a}_{3}}<\mathsf{\mathsf{a}_{1}<}%
\frac{1-\sqrt{5}}{2}\mathsf{\mathsf{a}_{3}}$.

\section{Acknowledgments}

The authors acknowledge Prof F. W. Hehl for discussions and helpful
suggestions. They also express their gratitude to CNPq-Brasil for the
invaluable financial help.

\end{document}